\documentclass{aa}  
\usepackage{xcolor}

\usepackage{graphicx}
\usepackage[varg]{txfonts}

\begin{document} 

\title{Bidirectional anisotropic solar energetic particle events observed by Solar Orbiter}

   \author{Zheyi Ding          \inst{1}
           \and
           Robert F. Wimmer-Schweingruber \inst{1}
           \and 
           Yu Chen \inst{2}           
           \and 
           Lingling Zhao \inst{2}
           \and
           Alexander Kollhoff \inst{1}
           \and
           Patrick Kühl \inst{1}
           \and 
           Liu Yang \inst{1}           
           \and
           Lars Berger \inst{1} 
           \and
           Verena Heidrich-Meisner  \inst{1} 
           \and
           Javier Rodriguez-Pacheco \inst{3}
           \and
           George C. Ho \inst{4}         
           \and
           Glenn M. Mason \inst{5}
           \and
           Gang Li \inst{6}
           \and
           Tomáš Formánek \inst{7,8}
           \and 
           Christopher J. Owen \inst{9}  
          }

   \institute{ 
             \inst{1} Institute of Experimental and Applied Physics, Kiel University, Leibnizstrasse 11, D-24118 Kiel, Germany   \email{ding@physik.uni-kiel.de}\\
             \inst{2}  Center for Space Plasma and Aeronomic Research (CSPAR), The University of Alabama in Huntsville, Huntsville, AL 35805, USA\\
             \inst{3} Universidad de Alcalá, Alcalá de Henares 28805, Spain \\   
             \inst{4} Southwest Research Institute, San Antonio, TX 78228, USA\\
             \inst{5} Johns Hopkins Applied Physics Lab, Laurel, MD 20723, USA \\       
             \inst{6} State Key Laboratory of Lunar and Planetary Sciences and CNSA Macau Center for Space Exploration and Science, 
Macau University of Science and Technology, Macau, China \\
             \inst{7} Czech Academy of Sciences, Institute of Atmospheric Physics, Department of Space Physics, Boční II 1401, Prague 4, Czechia\\
             \inst{8} LIRA, Observatoire de Paris, Université PSL, CNRS, Sorbonne Université, Université Paris-Cité, Meudon, France\\
             \inst{9}  Mullard Space Science Laboratory, University College London, Holmbury St Mary, Dorking, Surrey RH5 6NT, UK
             }

   \date{Received ; accepted }

 
\abstract
{Solar energetic particle (SEP) events are critical for understanding particle acceleration and transport in the heliosphere. While most SEP events involve outward streaming particles along open magnetic field lines, bidirectional events characterized by simultaneous sunward and anti-sunward particle flows offer unique insights into magnetic field topology and the interplay of multiple acceleration sources.}
{We investigate the origin and transport of energetic particles in two rare bidirectional anisotropic SEP events observed by Solar Orbiter, with a particular emphasis on their association with magnetic flux ropes.}
{Energetic particles, solar wind plasma, magnetic field, and solar radio measurements were analysed. Via the velocity dispersion analysis, we determined release times and path lengths for distinct particle populations. Automated flux rope identification and magnetic helicity diagnostics were used to characterize magnetic flux ropes.}
{Both events showed two clear velocity dispersion signatures with opposite particle anisotropies during their onset phase. The sunward streaming protons, characterized by a delayed release time, a harder spectral index, and higher intensities, may be due to coronal mass ejection-driven shock acceleration, while the promptly released anti-sunward streaming protons are likely linked to flare acceleration. Notably, in both cases, small-scale flux ropes were identified in situ during the time intervals corresponding to the bidirectional particle streaming. Path lengths derived for sunward and anti-sunward injections were substantially greater than nominal values of the Parker field lines, further evidence of the role of the flux rope in shaping particle trajectories. }
{These observations demonstrate that magnetic flux ropes can significantly affect magnetic connectivity to the source region and SEP propagation in the inner heliosphere, and that simultaneous velocity dispersion from two distinct particle sources can be used to place direct constraints on the topology of the flux rope. Our results highlight the value of combining particle anisotropy, release time, source spectra, and magnetic structure diagnostics to unravel SEP transport in complex transient magnetic structures, and also present new challenges for the current SEP transport model. }

   \keywords{solar wind – Sun: particle emission – Sun: magnetic fields – acceleration of particles – Sun: coronal mass ejections (CMEs)  }
   \titlerunning {Bidirectional SEP events}
   \authorrunning{Ding et al.}
   \maketitle
%
\nolinenumbers

\section{Introduction}
\label{sec:intro}

Solar energetic particle (SEP) events are characterized by transient enhancements in the intensity of energetic charged particles, which are accelerated from the solar corona during eruptive phenomena such as flares and coronal mass ejections (CMEs) and subsequently propagate along magnetic field lines into interplanetary space \citep[e.g.][]{Desai2016}. The particle anisotropy measured at the spacecraft serves as a direct indicator of particle transport processes. Specifically, a high anisotropy is typically associated with weak pitch angle scattering, meaning that the particle flux is strongly peaked along certain magnetic field directions, with most particles streaming either sunwards or anti-sunwards. In contrast, a low or near-zero anisotropy suggests strong pitch angle scattering or enhanced perpendicular diffusion, resulting in a more isotropic particle distribution \citep{Dresing2014A&A}. Therefore, anisotropy is a crucial diagnostic for identifying atypical SEP transport processes, especially in complex interplanetary environments.
 
While most SEP events exhibit outward streaming particles along open magnetic field lines, an increasing number of in situ observations have revealed rare cases of sunward-directed SEP events, where particles are detected propagating back towards the Sun with high intensity \cite[e.g.][]{Richardson1996JGR,Larson1997GeoRL,Rodriguez-Pacheco1999ICRC,Richardson2000JGR,Malandraki2002,rodriguez2003topology,leske_large_2012,Tan2012ApJ,Gomez-Herrero2017ApJ,Li2020ApJ,Wei2024ApJL,Rodriguez-Garcia2025A&A}. These events typically involve special interplanetary conditions, such as an observer being embedded within a magnetic cloud (MC), the presence of a shock or compressed magnetic fields acting as a magnetic mirror, or remote acceleration sites away from the observer.
Based on previous case studies, sunward SEP events can be generally categorized into four main types according to their underlying physical mechanisms. Firstly, the observer can reside within a closed magnetic flux rope (e.g. a MC) with both legs anchored at the Sun, enabling bidirectional (sunward and anti-sunward) particle flows via dual injections from two legs or reflection at the footpoint \citep{Dresing2016A&A, Gomez-Herrero2017ApJ, Rodriguez-Garcia2025A&A}. Moreover, the three-dimensional topology of the flux rope also allows for the possibility that particles are injected from the north or south directions \citep{wimmer2023unusually}.  Secondly, sunward SEPs can arise from magnetic mirroring, where SEPs move along open field lines and encounter a region of increased magnetic field strength, such as a shock or a CME sheath, which acts as a magnetic mirror and reflects particles towards the Sun \citep{Malandraki2002, Klassen2012A&A,Tan2012ApJ, Li2020ApJ}. 
In contrast to cases where an observer inside a closed flux rope can detect SEPs from different sources at each footpoint, the mirror effect results in the observation of particles originating from the same source.
Thirdly, \citet{Wei2024ApJL} propose that a remote strong accelerator (e.g. an interplanetary CME-driven shock) located beyond the observer can inject SEPs that then propagate sunwards along open magnetic field lines to the spacecraft. Finally, cross-field diffusion and field-line meandering may allow sunward SEPs to reach observers with poor magnetic connectivity to the source region \citep[e.g.][]{He2015ApJ}. However, this mechanism alone is insufficient to explain sunward-propagating events observed at locations with good magnetic connectivity to the source.

While previous studies have established several categories of sunward-propagating SEP events, here we investigate bidirectional SEP events that exhibit two distinct velocity dispersion signatures during their onset phases, one associated with anti-sunward propagation and the other with sunward propagation. Such events, although rare, can provide critical constraints on both the path lengths and release times of different particle sources and thereby offer unique insights into the large-scale configuration of the interplanetary magnetic field, as well as the particle acceleration and transport.
In this work we present a detailed analysis of two bidirectional, anisotropic SEP events observed by Solar Orbiter, each displaying two distinct velocity dispersion signatures and both likely associated with small-scale flux ropes (SFRs). To our knowledge, such events have not been previously reported. SFRs typically have durations ranging from tens of minutes to several hours. The origin of SFRs remains a subject of debate, as they are believed to result from solar eruptive processes, possibly as manifestations of small CMEs \citep{Feng2007JGRA}, and from in situ formation in the heliosphere through magnetic reconnection \citep{Cartwright2010JGRA}.
Notably, SFRs have much shorter durations and lack the typical plasma signatures of large-scale CMEs \citep[e.g.][]{Wimmer2006SSRv,Hu2018ApJS}.  As a result, SFRs are more challenging to detect and interpret, yet they could play an important role in modulating energetic particle propagation in the solar wind. 

Through a combination of energetic particle measurements, velocity dispersion analysis (VDA), and magnetic field diagnostics, we investigated the origins, transport process, and associated magnetic structure of two bidirectional SEP events. Section~\ref{sec:methods} describes the data and methods employed. The observations and interpretation of the two events are presented in Sect.~\ref{sec:results}. Finally, Sect.~\ref{sec:conclu} summarizes the main findings and discusses their implications for our understanding of SEP transport and the topology of flux ropes.

\section{Data and methods}\label{sec:methods}

This study utilizes data collected by multiple instruments on board Solar Orbiter. Specifically, SEP measurements are provided by the Energetic Particle Detector (EPD) suite (\citealt{Pacheco2020A&A...642A...7R, Wimmer-Schweingruber2021A&A...656A..22W}), which comprises the Supra-Thermal Electron and Proton sensor (STEP), the Electron Proton Telescope (EPT), the High Energy Telescope (HET), and the Suprathermal Ion Spectrograph (SIS). These instruments cover ion energies from a few keV nucleon$^{-1}$ up to above 100 MeV nucleon$^{-1}$. In this work we mainly utilized ion measurements from EPT and HET; each offers four viewing directions, thereby enabling the analysis of particle anisotropy.  While EPT does not explicitly differentiate between ion species, the measured fluxes are commonly interpreted as proton-dominated, since protons constitute the most abundant ion population in large SEP events.
Magnetic field data are obtained from the Solar Orbiter Magnetometer (MAG; \citealt{Horbury2020A&A...642A...9H}). Plasma parameters are measured with the Proton-Alpha Sensor (PAS), and the pitch angle distributions of suprathermal electrons are provided by the Electron Analyser System (EAS), both part of the Solar Wind Analyzer (SWA; \citealt{Owen2020A&A...642A..16O}). Additionally, we examined radio data from the Radio and Plasma Waves (RPW; \citealt{maksimovic2020solar}) instrument on board Solar Orbiter.

A key focus of this work is to investigate the roles of small-scale magnetic flux ropes in the bidirectional SEP events. Identifying SFRs in the solar wind presents a significant challenge due to their short durations, weaker coherent magnetic signatures compared to MCs, and ambiguous boundaries, especially in the presence of solar wind fluctuations. To address this, we employed the automated detection algorithm implemented in the PyGS package \citep{Hu2018ApJS,Zheng2018ApJL,Chen2022ApJ}, which leverages the unique characteristics of the Grad–Shafranov (GS) equation \citep{Sonnerup1996GeoRL, Hu2002JGRA}.  In the local flux rope frame, a spacecraft crossing an SFR typically observes a bipolar rotation in one of the magnetic field components, corresponding to nested isosurfaces of the magnetic flux function $A_{f}$. For a true flux rope, the transverse pressure ($P_t$) is expected to be a single-valued function of $A_{f}$, resulting in a characteristic double-folding pattern of $P_t(A_{f})$ between the inbound and outbound traversals. The PyGS algorithm systematically scans in situ data for intervals that exhibit this double-folding signature, enabling a robust and automated identification of SFRs.

However, in highly dynamic or turbulent environments, conventional automated methods such as PyGS may fail to reliably identify flux rope structures due to strong magnetic field fluctuations. To address these limitations, we supplemented our analysis with calculations of the normalized reduced magnetic helicity, $\sigma_m$, which serves as a sensitive proxy for magnetic field twist.
Following \cite{Zhao2020ApJS,Zhao2021A&A_detection},  the scale- and time-dependent normalized reduced magnetic helicity, $\sigma_m(s, t)$, is then evaluated as
\begin{equation}
\sigma_m(s, t) = \frac{2 \mathrm{Im}(\tilde{B}_T^* \tilde{B}_N)}{|\tilde{B}_R|^2 + |\tilde{B}_T|^2 + |\tilde{B}_N|^2},
\end{equation}
where the tilde represents wavelet-transformed quantities using the complex Morlet wavelet function \citep{torrence1998practical}, $\mathrm{Im}$ refers to the imaginary part, $s$ represents the wavelet scale, and the asterisk denotes complex conjugation. Here, $B_R$, $B_T$, and $B_N$ correspond to the magnetic field components in the radial-tangential-normal (RTN) coordinate system.
As demonstrated in previous studies \citep{Zhao2020ApJS, Zhao2021A&A_turbulence, Zhao2021A&A_detection}, flux ropes are typically characterized by high values of magnetic helicity, which result from the large-angle rotation of the magnetic field.  In this work we used regions with $|\sigma_m| > 0.6$ as proxies for the possible presence of SFRs. Since strongly Alfvénic fluctuations also typically appear as intermittent enhancements of $|\sigma_m|$, to exclude their influence, we further examined the normalized cross helicity $\sigma_c$ and the normalized residual energy $\sigma_r$, as described in Appendix~\ref{appendix:wavelet}.

\section{Results}\label{sec:results}

\begin{figure*}
\sidecaption
\centering
\makebox[\hsize][c]{\resizebox{0.75\hsize}{!}{\includegraphics{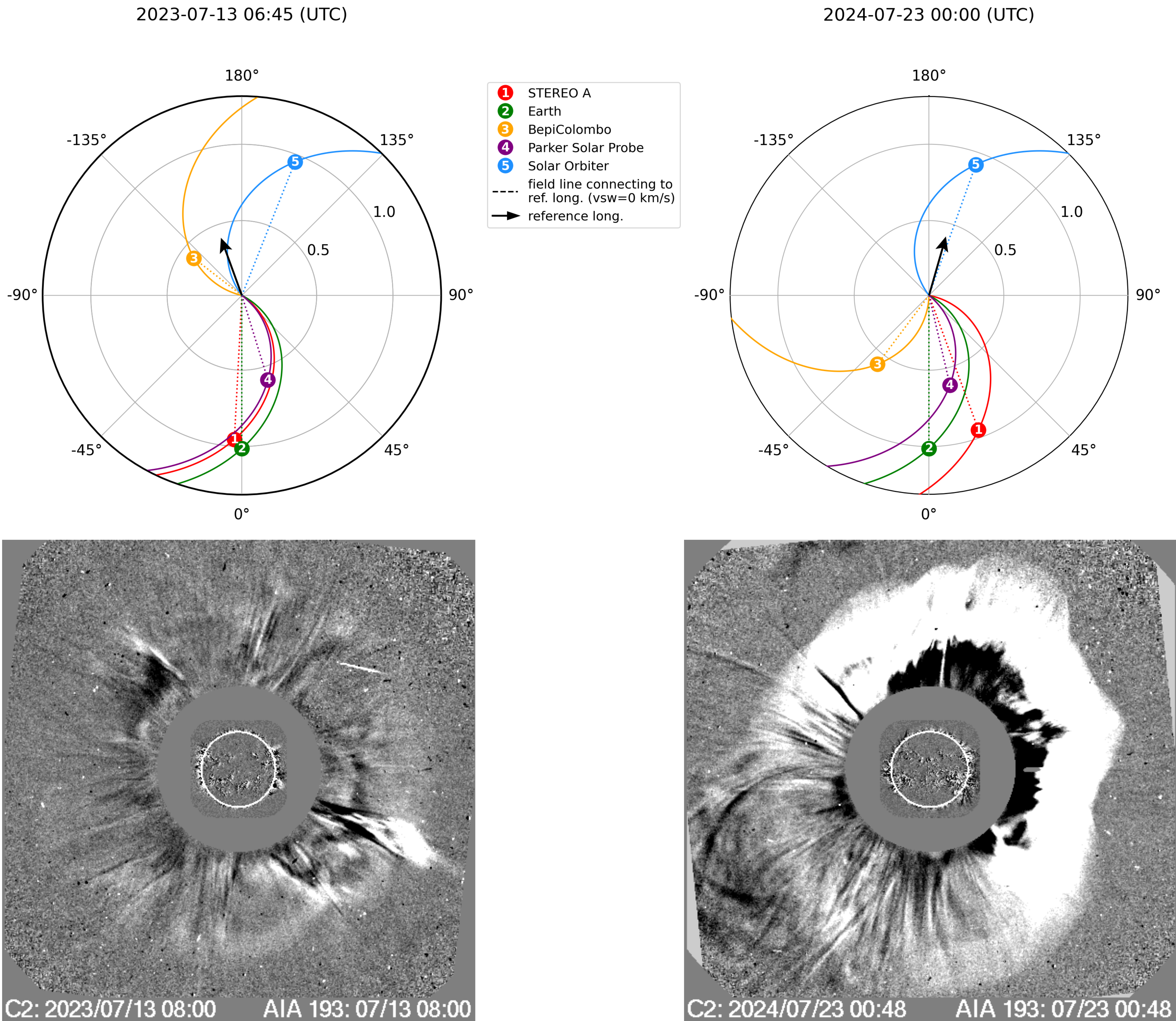}}}
\caption{Upper panels: Positions of the spacecraft during events 1 and 2 from Solar-MACH \citep{gieseler2023solar}. The location of Solar Orbiter is indicated by a blue dot. The arrow shows the direction of the associated solar flare. The Parker spiral magnetic field line corresponds to an assumed solar wind speed of $400~\mathrm{km/s}$.
Lower panels: Corresponding halo CMEs observed by the LASCO-C2 coronagraph on board SOHO for both events. }
\label{fig:sc_location}
\end{figure*}
For clarity and ease of comparison, all figures of observations in this section consistently present the July 13, 2023, event (event 1) in the left-hand panels and the July 23, 2024, event (event 2) in the right-hand panels.
Figure~\ref{fig:sc_location} illustrates the two solar eruptions and the respective spacecraft locations analysed in this study. During event 1, an active region was located on the far side of the Sun at N15E160 as viewed from Earth (upper-left panel), while Solar Orbiter was positioned at $158^\circ$ Stonyhurst longitude. At approximately 06:46~UT, this region produced an intense X-ray flare, recorded by the Spectrometer and Telescope for Imaging X-rays (STIX;\citealt{Krucker2020A&A}) on board Solar Orbiter, with a magnitude estimated to be X1 on the GOES scale. This eruption was accompanied by a halo CME observed by SOHO/LASCO (lower-left panel), with the associated CME propagating at approximately $1287\;$km/s at a heliocentric distance of about $19~R_\odot$, according to the DONKI database \footnote{https://kauai.ccmc.gsfc.nasa.gov/DONKI/view/CMEAnalysis/25934/2}. 
Similarly, for event 2, Solar Orbiter was located at $160^\circ$ Stonyhurst longitude. An active region was present at S01W164, also on the far side of the Sun and towards Solar Orbiter (upper-right panel). At around 23:56~UT on July 22, 2024, this region produced a powerful flare estimated to be X14 on the GOES scale. The corresponding halo CME (lower-right panel) had an estimated speed of approximately $1299\;$km/s at a heliocentric distance of about $19~R_\odot$ \footnote{https://kauai.ccmc.gsfc.nasa.gov/DONKI/view/CMEAnalysis/32125/2}.  We note that both events originated on the far side of the Sun, so the estimated CME speeds may be subject to significant uncertainty.

\begin{figure*}
\sidecaption
\centering
\makebox[\hsize][c]{\resizebox{1.0\hsize}{!}{\includegraphics{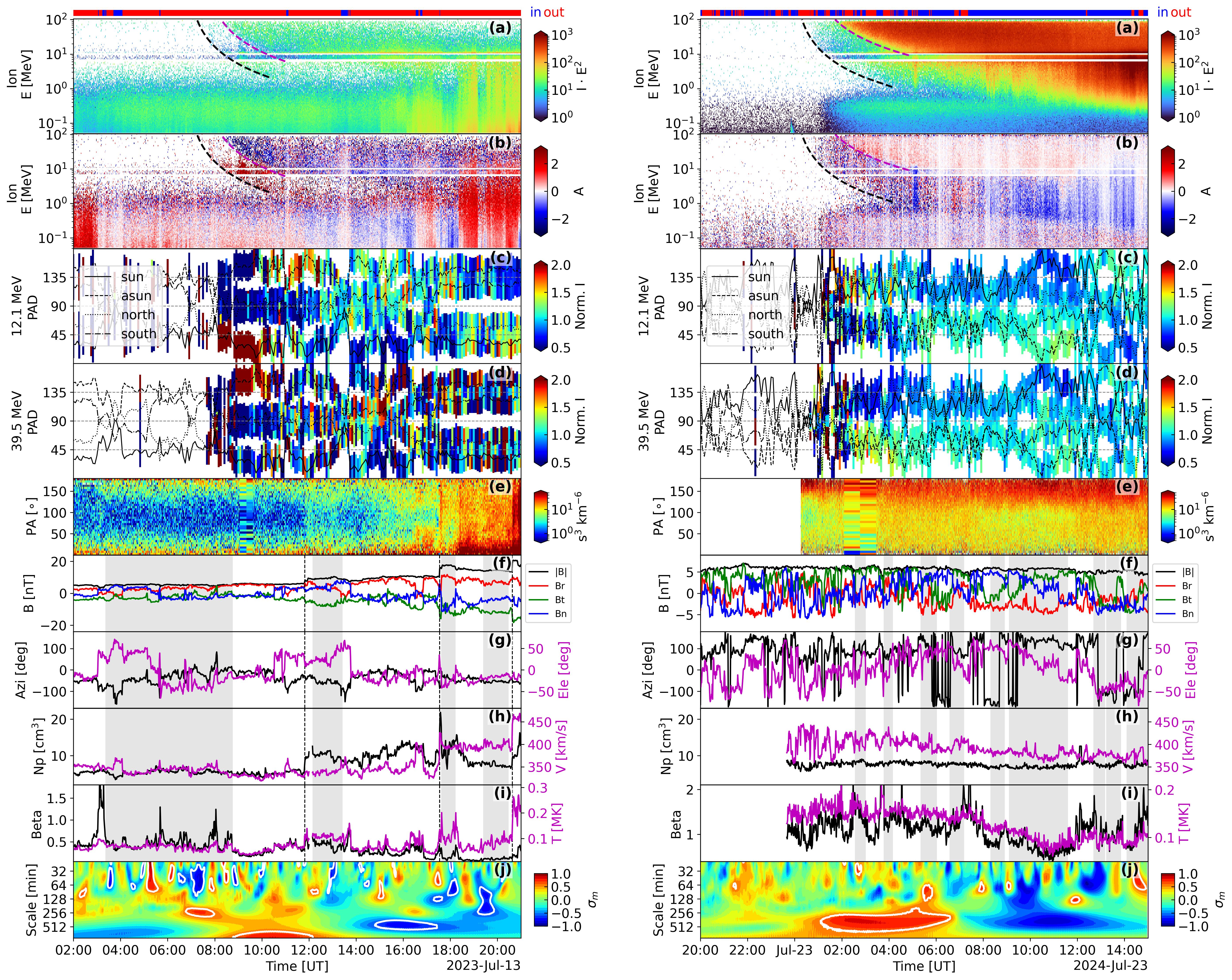}}}
\caption{Overview of SEP events 1 (left) and 2 (right). The top horizontal bars show the in situ magnetic field polarity, with red (blue) indicating a magnetic field direction outwards from (inwards towards) the Sun. Panels (a): Omnidirectional proton dynamic spectra measured by EPT and HET. The colour-coded bins represent scaled intensity $I \cdot E^2$ in units of $\mathrm{cm}^{-2}~\mathrm{s}^{-1}~\mathrm{sr}^{-1}~\mathrm{MeV}$. Panels (b): First-order anisotropy of protons. Panels (c): Pitch angle distribution (PAD) of $12.1~\mathrm{MeV}$ protons measured by the four HET telescopes. Panels (d): PAD of $39.5~\mathrm{MeV}$ protons. Panels (e): PAD of $>70~\mathrm{eV}$ suprathermal electrons measured by SWA/EAS. Panels (f): Magnetic field strength in RTN coordinates. Panels (g): Azimuthal and elevation angles of the magnetic field. Panels (h): Proton number density and bulk solar wind speed. Panels (i): Plasma beta and proton temperature. Panels (j): Normalized reduced magnetic helicity, with white contours indicating regions where $|\sigma_{m}| > 0.6$. The vertical dashed lines mark the locations of interplanetary shocks, and the grey-shaded area denotes the intervals of the SFRs.
}
\label{fig:event_overview}
\end{figure*}

\begin{figure*}
\sidecaption
\centering
\makebox[\hsize][c]{\resizebox{1.0\hsize}{!}{\includegraphics{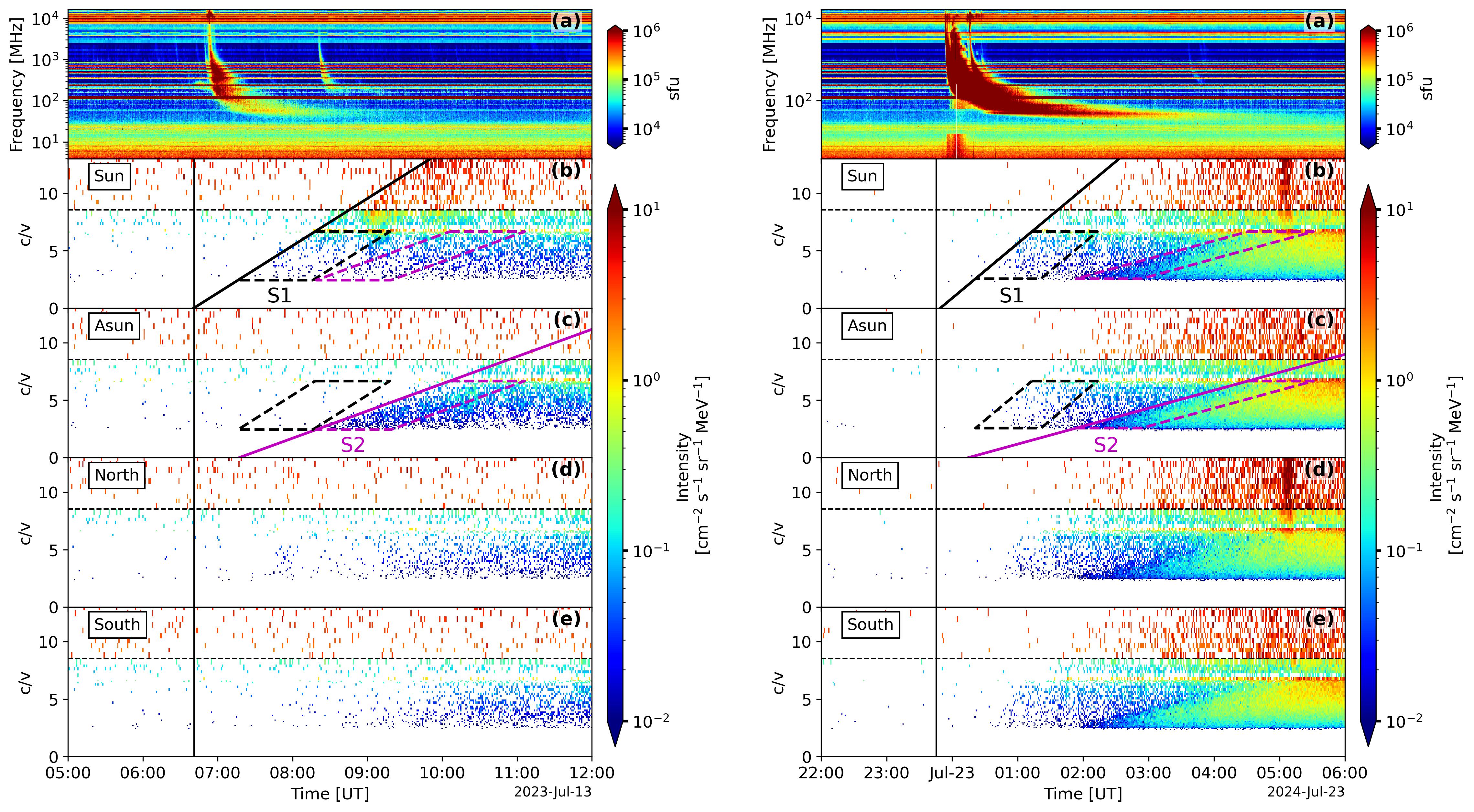}}}
\caption{Top panel: Radio dynamic spectrum observed by SolO/RPW. 
Lower four panels: Proton intensity as a function of inverse beta ($c/v$, where $c$ is the speed of light and $v$ is the particle velocity) and time for the sunward, anti-sunward, north, and south telescopes of both EPT and HET. The vertical line marks the onset time of the radio burst, which has been shifted $7.9$ and $7.2$ minutes earlier to facilitate comparison with the particle release time for event~1 and event~2. The horizontal dashed black  line in each panel marks the boundary between the energy ranges of HET (below) and EPT (above). In panels (b) and (c), the dashed black and magenta lines represent the fitted velocity dispersion tracks derived from the sunward and anti-sunward telescopes, respectively; the black and magenta boxes indicate the time intervals and energy ranges used for spectral integration within a 1-hour window (see Fig.~\ref{fig:spectra} for details). 
   }
\label{fig:vda}
\end{figure*}

Figure~\ref{fig:event_overview} provides an overview of the two SEP events examined in this study,  event~1 and event~2,  as observed by Solar Orbiter at heliocentric distances of $0.95~\mathrm{au}$ and $0.92~\mathrm{au}$, respectively. 
We first focus on event~1, as shown in the left-hand panels. The top bar indicates that the in situ magnetic field direction is predominantly outwards from the Sun during the interval of interest. Panel~(a) presents the omnidirectional proton dynamic spectra with a time resolution of $60~\mathrm{s}$, revealing an initial weaker particle injection followed by a much stronger injection spanning energies from a few MeV up to $100~\mathrm{MeV}$ (for the dynamic spectra of the individual viewing direction, see Fig.~\ref{fig:vda}). The black and magenta dashed lines indicate the fitted velocity dispersion tracks (more details are provided in Fig.~\ref{fig:vda}). 
The first-order anisotropy $A$, shown in panel~(b), is a key diagnostic in this work. We derived the first-order anisotropy using directional intensity measurements from both EPT and HET, which both measure in four directions.  The first-order anisotropy is defined as $A = (3 \int_{-1}^{+1} I(\mu)\cdot\mu d\mu)/(\int_{-1}^{+1} I(\mu) d\mu)$, where $I(\mu)$ is the pitch-angle-dependent intensity in a given direction and $\mu$ is the average pitch angle cosine. Here we computed the anisotropy for each time and energy bin. The sign of $A$ is interpreted in the context of the magnetic field polarity: for a positive (outward) field, positive $A$ indicates anti-sunward particle flow, while negative $A$ corresponds to sunward flow. The interpretation is reversed for negative polarity.
Given that the magnetic field is mostly outwards in event~1, large positive anisotropy indicates that particles are streaming away from the Sun, whereas negative anisotropy denotes sunward streaming.  A pronounced anisotropy (its magnitude close to 3) is observed at energies above $1~\mathrm{MeV}$, and its velocity dispersion suggests two separate injections within the first six hours following the event onset. 
Notably, the higher particle intensity corresponds to the sunward particle flow.
Panel~(c) displays the pitch angle distribution of $12.1~\mathrm{MeV}$ protons, with the colour scale representing normalized intensity, calculated as the ratio of the flux measured by each telescope to the mean flux across four telescopes.  After the onset, the intensity is initially highest for pitch angles below $45^\circ$ (from the sunward telescope) and then shifts to above $135^\circ$ (anti-sunward telescope), indicating a stronger particle flow directed towards the Sun later on. 
Panel~(d) shows the pitch angle distribution of $39.5~\mathrm{MeV}$ protons. During the period of 9:00 to 10:00~UT, the $39.5~\mathrm{MeV}$ protons show peak intensity with pitch angles $>135^\circ$, while the $12.1~\mathrm{MeV}$ protons display maximum intensity with pitch angles $<45^\circ$. This comparison demonstrates that Solar Orbiter observed two populations from opposite directions simultaneously. 
Panel~(e) presents the pitch angle distribution of $>70~\mathrm{eV}$ electron measurements from SWA/EAS. A distinct bidirectional electron streaming is apparent from around 03:00 to 12:00~UT on July 13, 2023, as indicated by enhanced phase space density near pitch angles $0^\circ$ and $180^\circ$. This suggests that Solar Orbiter might be situated on a closed magnetic structure with both footpoints connected to the Sun.
Based solely on the plasma and magnetic field measurements, it is not possible to visually confirm a classical magnetic flux rope between 03:00 and 12:00 UT due to the absence of clear rotations or significant plasma signatures from panels~(f)-(i).  To further characterize the possible magnetic structure, we applied the PyGS algorithm to identify possible SFRs during this interval. The resulting SFRs are indicated by the grey-shaded area, with a primary SFR detected between 03:21 and 08:45~UT.  This flux rope is likely responsible for the observed bidirectional streaming particle in this event, although its start and end times do not fully coincide with the period of bidirectional electrons. As discussed in Sect.~\ref{sec:intro}, the origin of SFRs can be either locally generated or associated with small CMEs. Given the relatively long duration of the SFR (approximately 5 hours) and the presence of bidirectional suprathermal electrons, we suggest that this structure likely corresponds to a small interplanetary CME. Panel~(j) displays the normalized reduced magnetic helicity, with white contour lines enclosing regions where $|\sigma_{m}| > 0.6$. These high magnetic-helicity regions are observed between around 05:00 and 08:00~UT, coinciding with the duration of the SFR identified by PyGS.  We further focused on this interval and examined the normalized cross helicity ($\sigma_c$) and normalized residual energy ($\sigma_r$), which suggest the presence of an Alfvénic SFR, as detailed in Appendix~\ref{appendix:wavelet}.  

We attempted to trace the origin of the SFR back to the Sun. According to the DONKI database, multiple small CMEs erupted from the same active region as event~1 between July 8, 2023, and July 10, 2023. This scenario is also supported by the identification of three in situ shocks on July 13, 2023 \citep{trotta2025overview}, as indicated by the vertical dashed lines in panels~(f)-(i). Assuming the speed of a small CME ranges from $350~\mathrm{km/s}$ to $400~\mathrm{km/s}$ as suggested by in situ solar wind measurements, the propagation time from the Sun to Solar Orbiter is approximately $4.1$ to $4.7$ days 
This implies an eruption time between July 8, 2023, at 10:00~UT and July 9, at 00:00~UT, given an in situ onset of SFR at around 03:00~UT on July 13, 2023. Within this window, only one candidate CME was recorded, launched on July 8, 2023, at 20:36~UT, with a reported speed of $415~\mathrm{km/s}$ and a half-width of $32^\circ$ \footnote{https://kauai.ccmc.gsfc.nasa.gov/DONKI/view/CME/25858/1}. The associated flare was located at N15W140 as viewed from Earth. 
Therefore, the SFR and event~1 likely originated from the same active region. However, we did not investigate this further due to limited observational evidence.

Compared to event 1, event 2 exhibits more complex features, as shown in the right-hand panels of Fig.~\ref{fig:event_overview}. There are data gaps in the SWA measurements during the pre-event period on July 22, 2024. The proton dynamic spectra in panel~(a) also display two distinct tracks of velocity dispersion. Bidirectional streaming characteristics are again apparent in the first-order anisotropy shown in panel~(b). In this event, the magnetic field polarity is predominantly inwards, so negative first-order anisotropy indicates particles propagating away from the Sun, whereas later positive values correspond to sunward streaming. This interpretation is reflected in panel~(c), where the highest intensity of $12.1~\mathrm{MeV}$ protons initially appears at pitch angles around $135^\circ$ (as measured by the sunward telescope) but later shifts to around $45^\circ$ (anti-sunward telescope). Panels~(c) and (d) reveal that pitch angle distribution in this event exhibits more pronounced fluctuations than in event~1. 
The pitch angle distribution of $39.5~\mathrm{MeV}$ protons in panel~(d) shows a feature that is similar to event~1, during the period of 2:00 to 4:00~UT, the highest intensity of $39.5~\mathrm{MeV}$ protons is measured with pitch angle around $45^\circ$ but the highest intensity of $12.1~\mathrm{MeV}$ protons appears at pitch angles around $135^\circ$, indicating the bidirectional SEP propagation.
Unlike event~1, the bidirectional electron feature measured by EAS is less evident in the pitch angle distribution shown in panel~(e). Although the phase space densities near both $0^\circ$ and $180^\circ$ are elevated compared to $90^\circ$, the peak at $180^\circ$ is more pronounced in the inward magnetic field configuration. The magnetic field and solar wind data in panel~(f)-(i) reveal substantial fluctuations, making it challenging to identify SFRs directly using PyGS. Only a few short-duration and unrelated SFRs are detected. 
Notably, analysis of the normalized reduced magnetic helicity reveals the presence of a highly helical magnetic structure with a characteristic scale of approximately $512$ minutes, spanning from around 00:00 to 06:00~UT on July 23, 2024. Although the precise boundaries are difficult to determine using the wavelet method, the combined results of $\sigma_c$ and $\sigma_r$ in Appendix~\ref{appendix:wavelet} indicate that Solar Orbiter crossed a potentially Alfvénic SFR embedded within a turbulent solar wind environment near the onset of the eruption. Unlike event 1, where the SFR was encountered before the eruption, in event 2, the encounter with the SFR is unclear. It seems to occur during the onset phase, which could allow for the observation of bidirectional SEPs.
Recent studies have emphasized the prevalence of such SFRs in the slow solar wind \citep[e.g.][]{chen2021small,shi2021parker}, where they exhibit a high degree of Alfvénicity. The possible Alfvénic flux ropes may have a significant impact on SEP transport by affecting both the propagation path and scattering conditions. Unfortunately, we did not find any relevant eruption information associated with this possible SFR in the DONKI database. If the small eruption occurred precisely on the far side of the Sun, it is possible that no evidence would be detectable in coronagraph observations near Earth.

To further understand the transport processes in the two events, we employed VDA to determine the release times and path lengths of the two sources. In the left-hand panels of Fig.~\ref{fig:vda}, the upper panel displays the type III radio burst observed by RPW during event~1. Only a single burst is evident during the onset phase, beginning at approximately 06:41~UT from the Sun, indicated by the vertical line. The lower four panels show proton dynamic spectra from the four HET and EPT telescopes, highlighting the features of velocity dispersion.  Two distinct velocity dispersion tracks can be readily identified by visual inspection in both the sunward and anti-sunward telescopes. Notably, the earliest arriving particles detected by the sunward telescopes are scarcely observed by the anti-sunward telescope during the onset. In addition, only a small number of particles are detected in the north and south telescopes at pitch angles near $90^\circ$, indicating very limited scattering during the transport process. Compared to the proton measurements, velocity dispersion features are less pronounced in the electron data, likely due to high-level background noise. Therefore, we present the electron measurements by EPT in the Appendix~\ref{appendix:eletron} and do not perform VDA for electrons in this study.

Onset times are determined independently for the sunward and anti-sunward telescopes using the Poisson-CUSUM method \citep{lucas1985counted}. VDA yields an injection time for source 1 (S1) of July 13, 2023, 06:40~UT $\pm$ 4 minutes, and for source 2 (S2) of July 13, 2023, 07:17~UT $\pm$ 6 minutes (see the solid lines in panels~b and c). The release time of S1 coincides with the onset time of a type III radio burst. The derived path lengths are $1.75~\mathrm{au}$  $\pm$ $0.05~\mathrm{au}$  for S1 and $3.04~\mathrm{au}$  $\pm$ $0.15~\mathrm{au}$  for S2, indicating distinctly different magnetic connectivity for the two sources. 
For event 2, as shown in the right-hand panel of Fig.~\ref{fig:vda}, a similar analysis is performed. Onset time for the sunward telescope is determined using the Poisson-CUSUM method, while for the anti-sunward telescope it is estimated visually (by averaging the fit results from ten independent visual examinations), due to the high background intensity resulting from the first injection in this event. The VDA results show injection times of July 22, 2024, 23:45~UT $\pm$ 6 minutes for source~1 and July 23, 2024, 00:15~UT $\pm$ 8 minutes for source~2. The corresponding path lengths are $1.51~\mathrm{au}$  $\pm$ $0.08~\mathrm{au}$  for S1 and $4.62~\mathrm{au}$ $\pm$ $0.21~\mathrm{au}$  for S2. Due to the strong scattering, the path length for this event might be less robust, especially for the S2. Based on these results,  the derived onset times of S1 are reasonably consistent with the radio burst timing on July 22, 2024, at around 23:45 UT and with around $30$ minutes delay for S2.  
For both events, the path lengths derived for sunward and anti-sunward particles are significantly greater than the nominal Parker spiral path length at $0.95~\mathrm{au}$ for event~1 and at $0.92~\mathrm{au}$  for event~2.
By combining the estimated path lengths with the magnetic helicity of the identified flux rope, it is possible to infer the topology and spatial extent of the flux rope encountered in situ. An example of the calculation for event~1 is provided in Appendix~\ref{appendix:sfr}.

\begin{figure*}
\sidecaption
\centering
\makebox[\hsize][c]{\resizebox{1.0\hsize}{!}{\includegraphics{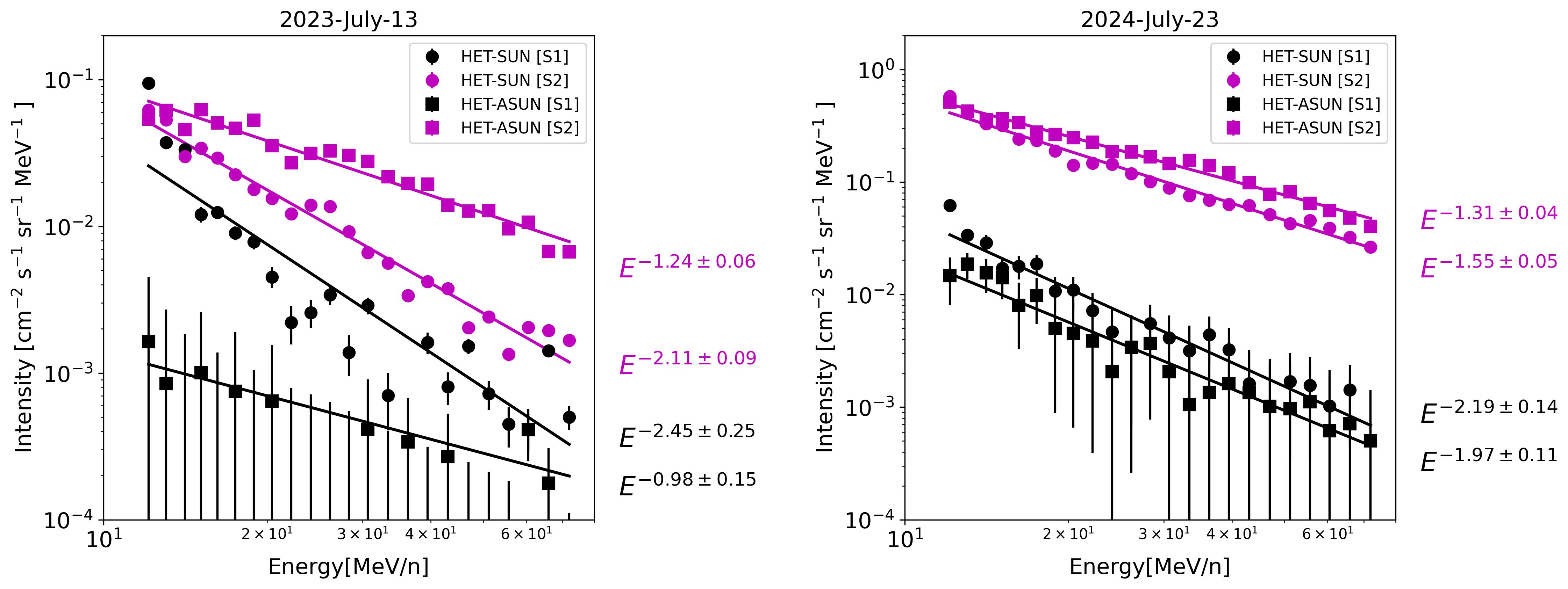}}}
\caption{ Time-averaged proton spectra from the HET sunward and anti-sunward telescopes. The black and magenta points correspond to the boxes of source~1 (S1) and source~2 (S2) in Fig.~\ref{fig:vda}, respectively. The lines represent single power-law fits, with the fitted spectral indices indicated on the right.}
\label{fig:spectra}
\end{figure*}

The different release times indicate possible different origins for the source particles for both events. The close temporal coincidence between the release time determined from the sunward telescope and the onset of the type III radio burst suggests that the anti-sunward streaming particles might be promptly accelerated and released during flare-related magnetic reconnection. In contrast, a release time delayed by approximately $30$ minutes beyond the onset of the type III radio burst, accompanied by higher particle intensities as observed from the anti-sunward telescope, may indicate that particles originate from CME-driven shock acceleration.
Numerous previous studies have shown that shock acceleration typically becomes efficient only when the CME-driven shock reaches high altitudes in the corona \citep[e.g.][]{Kahler1994ApJ,Cliver2004ApJ,Reames2009ApJ,Rouillard2011ApJ,Rouillard2012ApJ}.  This can be explained by two main processes. First, the formation of the shock itself: as the CME undergoes acceleration in the lower corona and exceeds the local fast magnetosonic speed, it drives the formation of a shock.  Second, once the shock forms, particle acceleration generally proceeds via diffusive shock acceleration, where the acceleration time of high-energy particles depends on shock properties and wave intensity near the shock. Both factors likely contribute to the late release of high-energy particles. Several studies have investigated the shock formation heights and release heights of high-energy SEPs to better constrain the coronal acceleration environment. \citet{gopalswamy2013height} determined the shock formation heights at the onset time of metric type II bursts for 32 SEP events between 2010 and 2012, finding shock heights ranging from $1.20~R_\odot$ to $1.93~R_\odot$ with a mean value of $1.43~R_\odot$.  \citet{Kouloumvakos2015A&A} analysed 65 SEP events and estimated the proton release heights to range from $1~R_\odot$ to $8~R_\odot$, with an average of $3.4~R_\odot$. More recently, \citet{Ameri2024SoPh} examined 58 high-energy proton events and found that events with $E_{\rm max}<100~\mathrm{MeV}$ typically exhibited proton release heights above $3.0~R_\odot$, suggesting that less energetic SEP events are more likely associated with acceleration processes occurring higher in the corona. Therefore, 
the delayed release time of sunward particles found in this work likely reflects the time required for the CME-driven shock to develop and become sufficiently strong to accelerate high-energy protons and release them onto magnetic field lines.

 In applying the traditional VDA, obtaining the onset time is often the most challenging step. For our events, we employed both the Poisson-CUSUM method and visual inspection to estimate the onset times, which inevitably introduces some degree of uncertainty. Nevertheless, the solid black line and purple lines in Fig.~3 (panels b and c) provide reasonable proxies for the onset times. Another issue with VDA is the assumption that particles of different energies are released from the source at the same time.    
Studies of energetic electrons in impulsive events suggest that such an assumption is reasonable with an uncertainty of $\sim 5$ minutes \citep{Zhao2019, wu_statistical2023}. For our events, we considered ions rather than electrons.  For both events,  if the first episode (S1) is due to flare acceleration,   we expect the uncertainty obtained in the release time to be around a few minutes.  For the second episode (S2), if the acceleration process is due to shock acceleration, then compared to that at flares,  the acceleration could take more time. This effect is even more pronounced for higher-energy ions. The net effect is that the release time obtained will have 
an uncertainty in the order of  the acceleration timescale ($t_{\mathrm{acc}}$) of the maximum proton energy $E_{\rm max}$ used in the VDA analysis. 
 The maximum energy achievable at the shock front is determined by equating the particle acceleration timescale ($t_{\mathrm{acc}} \sim \kappa/u^2$) with the shock  dynamic timescale \citep[e.g.][]{Li2017ScChD..60.1440L}, where $\kappa$ is the particle diffusion coefficient and $u$ is the upstream plasma speed in the shock frame. As the shock evolves quickly near the Sun,  the shock dynamic timescale quickly drops, so is the maximum particle energy \citep{Li+2003,Li2017ScChD..60.1440L}.  The maximum particle energy considered in both events is approximately $70$~MeV.  While direct measurements of the shock formation height and its evolution are not clear for the two events, we adopted representative values based on previous statistical studies as discussed above. We assumed the shock forms at $\sim 1.43~R_\odot$ and propagates outwards to a release height of $\sim 3~R_\odot$, maintaining relatively stable shock characteristics (compression ratio, shock speed, and turbulence level upstream of the shock) during this interval. Using an average shock speed of $750~\mathrm{km/s}$ for both events, we estimate the available acceleration time to be approximately $24$ minutes.
 Since the fitted release time of the second episodes trail those of the first episodes in both events by $\sim 30$ minutes, it is therefore plausible that the two episodes correspond to the same `eruption' process near the Sun, with the first episodes having shorter acceleration time, consisting with flare acceleration; and the second episodes having longer acceleration time, consisting with shock acceleration. 
  To support this, we considered the spectrum of both episodes (see below) as the flare acceleration and shock acceleration likely yield different spectral indices.

To further characterize the potential differences between source~1 and source~2 in both events, Fig.~\ref{fig:spectra} presents the time-averaged proton energy spectra from the sunward and anti-sunward telescopes, calculated within a one-hour window following the onset times indicated by the dashed black and magenta boxes for S1 and S2 in Fig.~\ref{fig:vda}.  The duration of one hour is chosen to increase statistical significance and to avoid overlap between the windows of S1 and S2. We note that shortening the time window does not significantly change the spectral features.
For clarity, the spectra from the north and south telescopes are omitted. We focused only on proton energies above $10~\mathrm{MeV}$ with clearly different anisotropies because the low-energy particle intensity observed in the anti-sunward telescope may be contaminated by a pre-existing anti-sunward streaming particle population. The spectral index at the high-energy end is often used as an indicator of the acceleration efficiency, although it can also be influenced by transport and escape processes \citep[e.g.][]{mewaldt2012energy,Desai2016ions}.   

In event~1, as shown in the left-hand panel of Fig.~\ref{fig:spectra}, the proton intensity from S1 (black) is predominantly observed by the sunward telescope, and exhibits a power-law distribution of $E^{-2.45}$. Although the anti-sunward telescope detects only a very low intensity, it still exceeds the pre-event background level.
For S2 (magenta), however, the particle intensity is higher in the anti-sunward telescope across all energy channels. Notably, the spectral index of source~2 in the anti-sunward telescope is harder (approximately $-1.24$) than that of source~1 in the sunward telescope (approximately $-2.45$), which is close to the limit of theoretical spectral index $-1$ for a high compression ratio ($s=4$) in diffusive shock acceleration theory. CME-driven shock acceleration is generally more efficient at producing higher SEP intensity compared to flare acceleration \citep[e.g.][]{Li+2005mixed,reames2013two}. The significantly higher particle intensity and harder spectral index are further evidence that the particles from S2 were accelerated by the CME-driven shock.

In contrast, in the right-hand panel of Fig.~\ref{fig:spectra}, event 2 exhibits much stronger scattering, as evidenced by the detection of particles in all telescopes in Fig.~\ref{fig:vda}. This indicates a more isotropic distribution compared to event 1 and results in nearly identical spectral indices for the sunward and anti-sunward telescopes.
Nevertheless, distinctions between the two sources remain evident. For S1, the particle intensity is approximately two orders of magnitude lower than that of S2, and the spectral index is softer, around $-2$. During the first injection (S1), the intensity observed by the sunward telescope is roughly twice that of the anti-sunward telescope. Conversely, during the second injection (S2), the intensity at the anti-sunward telescope is about twice that at the sunward telescope. These comparisons also support the conclusion that particles from the second injection predominantly stream towards the Sun and originate from the shock.

\begin{figure}
\sidecaption
\centering
\makebox[\hsize][c]{\resizebox{1.0\hsize}{!}{\includegraphics{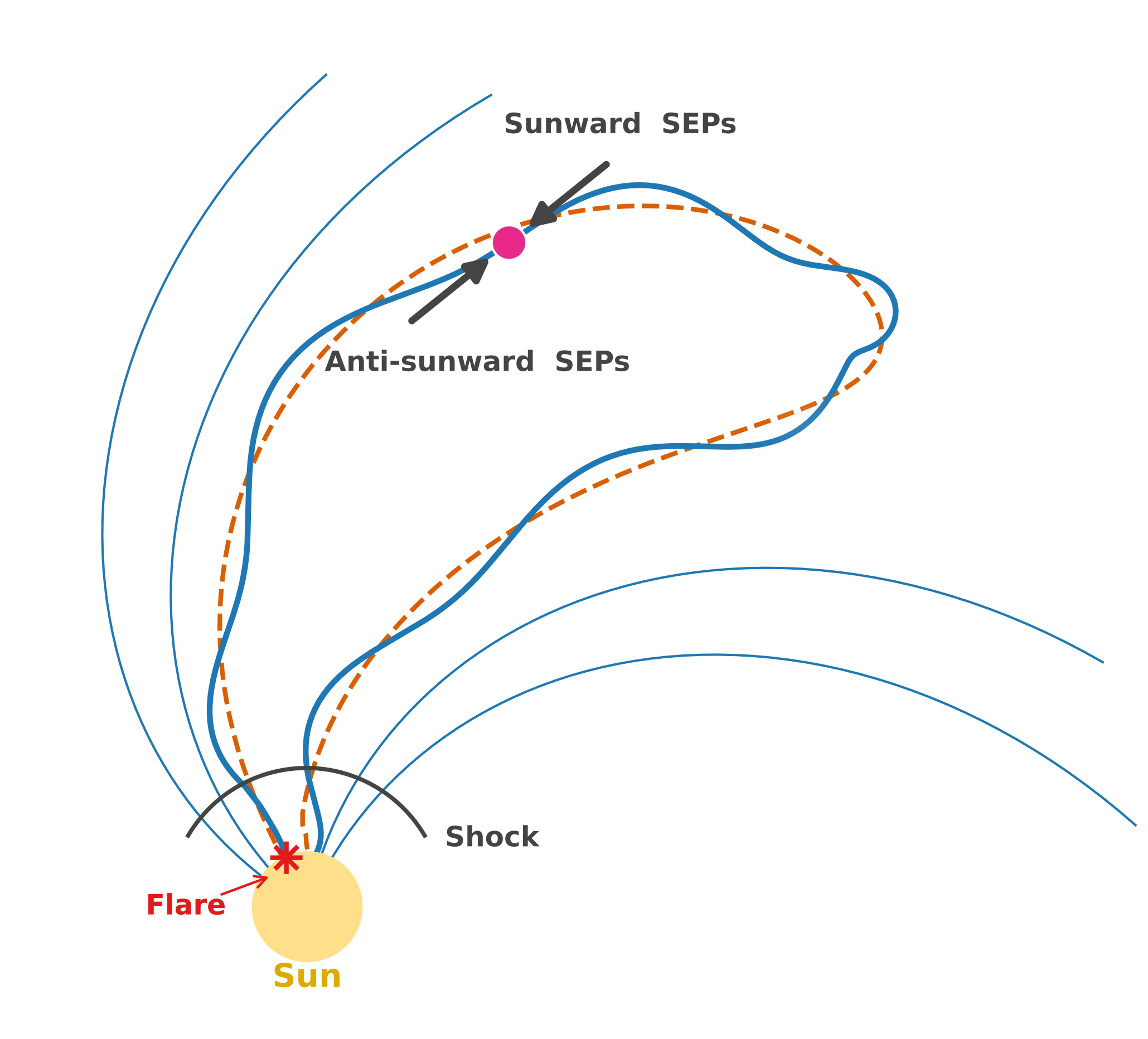}}}
\caption{ Schematic illustration of bidirectional SEP events within a magnetic flux rope structure. The observer (magenta circle) is embedded inside the flux rope (solid blue line), which is rooted at the Sun with both legs. The black arrows indicate the directions of SEP propagation associated with different possible sources: the anti-sunward SEPs primarily associated with flare-acceleration (red asterisk) and the sunward SEPs associated with particles accelerated by the CME-driven shock (black arc). }
\label{fig:flux_rope_schematic}
\end{figure}

Figure~\ref{fig:flux_rope_schematic} presents a two-dimensional schematic illustration of the possible origin of bidirectional SEPs within a magnetic flux rope, as discussed in this study. In this scenario, the observer is situated inside a flux rope rooted at the Sun, with both legs anchored in the lower corona and one leg connecting directly to the flare site. Following the eruption, anti-sunward SEPs originating from flare acceleration reach the observer first via a shorter path length, while sunward SEPs, associated with stronger CME-driven shock acceleration, arrive later after travelling a longer path.  
Particles accelerated at the shock front can also propagate along field lines connecting to the flare site later on. However, to observe a distinct bidirectional velocity dispersion, the particle intensity along the field line connecting to the sunward telescope should be lower than that connecting to the anti-sunward telescope. This difference may reflect the distinct shock geometries encountered by each leg. For example, quasi-parallel shocks may have a higher injection efficiency of seed particles compared to quasi-perpendicular shocks \citep{Zank2006,Li+2012,Ding2023JGRA..12831502D}. A detailed assessment of the shock geometry of these two events would require comprehensive shock modelling \citep[e.g.][]{Li2021,ding2022A&A...668A..71D,Ding2025A&A}, which is beyond the scope of this study but could be addressed in future work.
We note that the observer could also be located at the opposite leg of the flux rope, with this leg connected to the flare site. Due to the lack of multi-spacecraft observations, we do not further discuss the possible orientation or expansion direction of the flux rope. 
This configuration demonstrates how a flux rope enables the simultaneous observation of SEPs from different sources, characterized by distinct release times and path lengths, as discussed in our results.

\section{Discussions and conclusion}\label{sec:conclu}
In this work we have presented detailed analyses of rare bidirectional SEP events with two clear velocity dispersions during the onset phase, observed by Solar Orbiter on July 13, 2023, and July 23, 2024.
Both events exhibit two distinct velocity dispersion signatures with opposite anisotropies during their onsets, enabling unique constraints on the path lengths and release times of sunward and anti-sunward energetic particles. 
Based on observed differences in release times, path lengths, particle intensities, and spectral indices between the sunward and anti-sunward streaming populations, together with in situ identification of SFRs, we suggest that the anti-sunward streaming particles may originate primarily from solar flare acceleration, while the sunward streaming population could be associated with shock acceleration. In typical SEP events, flare-accelerated and shock-accelerated particles often propagate along the same interplanetary magnetic field lines to the observer, making it challenging to distinguish their contributions. However, for a spacecraft located within SFRs, the unique magnetic configuration provides an opportunity to separate the two sources because of their different path lengths. 

In turn, estimated path lengths for sunward- and anti-sunward-streaming particles provide a valuable means to constrain the topology of magnetic flux ropes. In both events, the presence of two possible distinct particle sources suggests a scenario in which both legs of the SFR remain magnetically connected to the Sun. Consequently, the central axial length of the SFR could be inferred by combining the measured particle path lengths with the estimated magnetic helicity of the flux rope, as discussed in Appendix~\ref{appendix:sfr}. However, in situ observations from a single spacecraft limit the ability to unambiguously determine the global structure of the SFR. Coordinated multi-spacecraft observations are required to place more stringent constraints on the topology of these structures.

Although the two events share a similar potential explanation, notable differences in local observational features are evident. Event 1 exhibits clear bidirectional particle flows with minimal scattering, leading to strong anisotropies and distinct velocity dispersion signatures for both sunward and anti-sunward particle populations. Moreover, the SFR structure associated with the onset of event 1 is identified by the PyGS algorithm and supported by bidirectional suprathermal electrons. In contrast, event 2 occurs within a more turbulent magnetic environment. Consequently, the SFR in event 2 is not reliably detected by PyGS and can only be inferred from wavelet analysis of the magnetic field. Additionally, bidirectional electron signatures in event 2 are less pronounced compared to event 1.  
Given the possible explanations discussed above, the associated SFRs in both events likely originated from the Sun and are probably linked to small CMEs. Although their spatial scales are smaller compared to large-scale flux ropes observed within MCs, the underlying magnetic topology may be similar. In both events, the particle propagation is governed by the closed nature of the flux rope, enabling sunward-streaming SEPs to be observed at Solar Orbiter. The identification of two velocity dispersions within the SFR intervals further supports a flux rope geometry that guides particles over extended distances along curved magnetic field lines, analogous to those in large MC events \citep[e.g.][]{leske_large_2012,Gomez-Herrero2017ApJ}.
The frequent occurrence of SFRs \citep{Hu2018ApJS,Chen2022ApJ}, some of which may originate from the Sun, suggests that a significant number of SEP events could be influenced by these elusive structures, which play a role in modulating particle transport.

Bidirectional SEP events with distinct velocity dispersion can also be explained by the mirror effect. For example, \cite{Li2020ApJ} report an SEP event at $1~\mathrm{au}$ where two electron populations were observed streaming away from and towards the Sun with similar release times but different path lengths. They suggest that this was due to reflection occurring beyond $1~\mathrm{au}$, rather than transport along a closed magnetic field topology. Under the mirror scenario, the sunward and anti-sunward particle populations are expected to exhibit comparable release times, particle intensities, and spectral indices. In contrast, our study finds large discrepancies in intensity and spectral slope between the two populations, with the sunward source significantly stronger, making the mirroring explanation unlikely in this case.  This also indicates that the clear detection of bidirectional SEP events associated with flux ropes requires more specific conditions. Most importantly, in a scattering environment, the first injection must be much weaker than the second. Otherwise, the initial particle population can dominate and mask the velocity dispersion and anisotropy signatures of the second injection. Such intensity differences might be more significant at higher energies due to varying acceleration efficiencies at different sites, which might explain why sunward SEPs are primarily observed in the HET energy channels.

In summary, our observations demonstrate that SFRs might be involved in shaping the transport and distribution of SEPs in the inner heliosphere. The bidirectional SEP events presented here provide potential evidence of an interplay between injections of different source particles and transient magnetic structures. Continued multi-spacecraft and high-resolution observations, combined with improved modelling efforts, will be essential to further unravel the topology of magnetic flux ropes and the particle transport process in bidirectional SEP events. 

\begin{acknowledgements}
     Solar Orbiter is a mission of international cooperation between ESA and NASA, operated by ESA. This work was supported by the German Federal Ministry for Economic Affairs and Energy and the German Space Agency (Deutsches Zentrum für Luft- und Raumfahrt, e.V., (DLR)), grant number 50OT2002. The UAH team acknowledges the financial support by Project PID2023-150952OB-I00 funded by MICIU/AEI/10.13039/501100011033 and by FEDER, UE.  The Suprathermal Ion Spectrograph (SIS) is a European facility instrument funded by ESA under contract number SOL.ASTR.CON.00004. Solar Orbiter post-launch work at JHU/APL is supported by NASA contract NNN06AA01C, at the Southwest Research Institute by NASA 80GFSC25CA035. Solar Orbiter Solar Wind Analyser (SWA) data are derived from scientific sensors which have been designed and created, and are operated under funding provided in numerous contracts from the UK Space Agency (UKSA), the UK Science and Technology Facilities Council (STFC), the Agenzia Spaziale Italiana (ASI), the Centre National d’Etudes Spatiales (CNES, France), the Centre National de la Recherche Scientifique (CNRS, France), the Czech contribution to the ESA PRODEX programme and NASA. Solar Orbiter SWA work at UCL/MSSL is currently funded under STFC/UKSA grants UKRI919 and UKRI1204. Y.C. acknowledges NASA grant 80NSSC21K1763 and NSF Grant AGS‐2229065 for support. L.Y. is partially supported by DFG under grant HE 9270/1-1. GL acknowledges 
    the Science and Technology Development Fund (FDCT) of Macau (grant Nos. 002/2024/SKL, 0008/2024/AKP).  T.F. acknowledges the funding by the France 2030 program, reference ANR-23-CMAS-0041, and the GAČR grant 22-10775S.
     This research was supported in part through high-performance computing resources available at the Kiel University Computing Centre. A Python package, PyGS, developed by Dr. Yu Chen for performing the GS reconstruction, is publicly available at https://github.com/PyGSDR/PyGS/.
\end{acknowledgements}

\bibliographystyle{aa} 
\bibliography{ref} 

\begin{appendix}
\onecolumn
\section{Supplementary figures}

\subsection{Directional electron measurements from EPT}
\label{appendix:eletron}

\begin{figure*}[h!]       
  \sidecaption              
  \centering
  \resizebox{\textwidth}{!}{\includegraphics{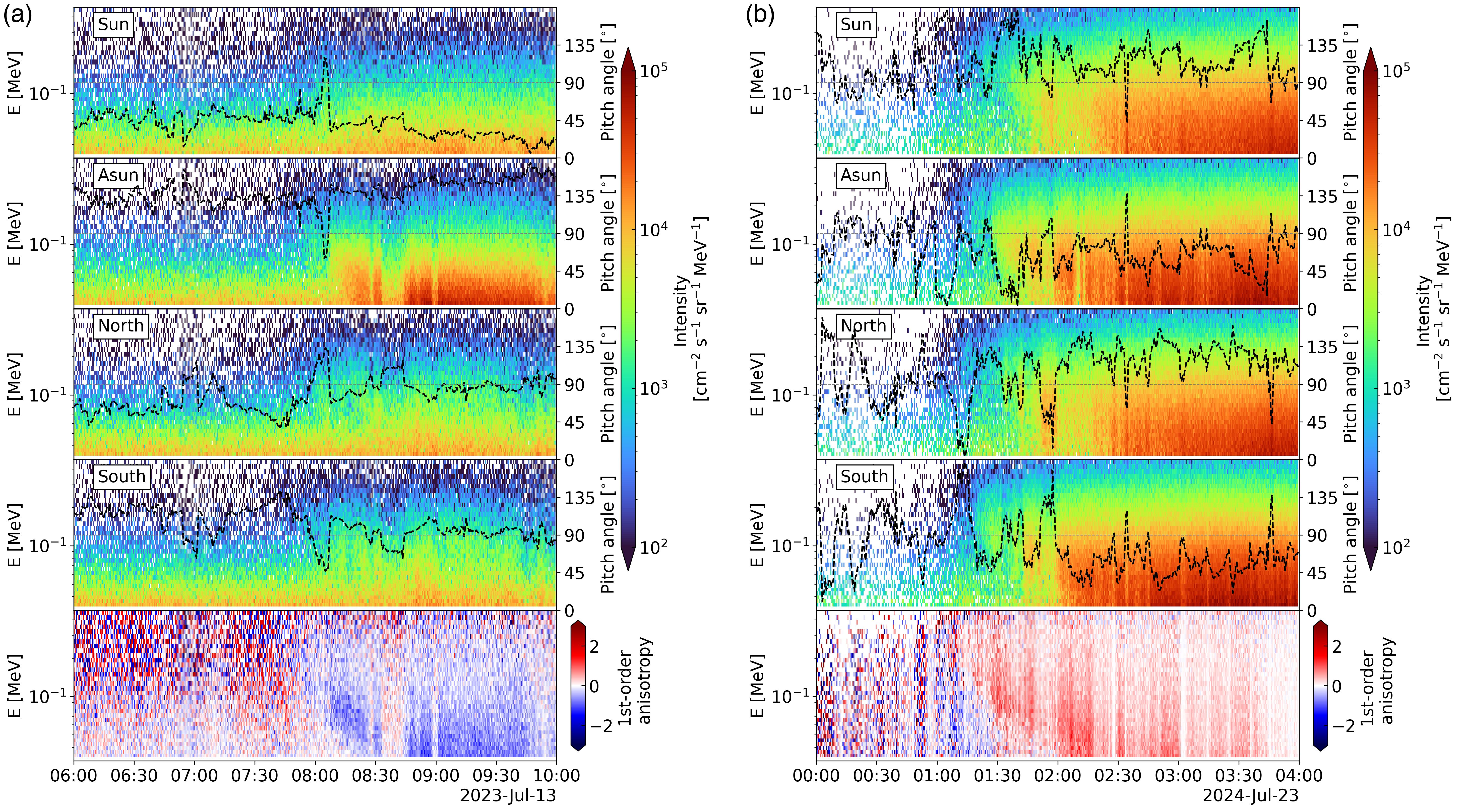}}
\caption{ Electron dynamic spectra for event 1 (left)  and event 2 (right) measured by the sunward, anti-sunward, north, and south telescopes of Solar Orbiter/EPT. }
    \label{fig:EPT_electron}
\end{figure*}

Figure~\ref{fig:EPT_electron} presents the colour-coded electron intensities measured by the sunward, anti-sunward, north, and south telescopes of Solar Orbiter/EPT for events 1 and 2. A comparison of the intensities among different telescopes reveals a pronounced anisotropy in the electron distribution. Analysis of the first-order anisotropy indicates the presence of bidirectional electron flows, with a majority of electrons streaming towards the Sun. The intensity of anti-sunward electrons is less pronounced, likely due to high background levels, which limit the ability to perform precise VDA. 

\subsection{Identification of magnetic flux ropes}
\label{appendix:wavelet}
To distinguish SFRs from Alfvén waves, the normalized magnetic helicity $\sigma_m$ alone is not sufficient.
Following \citet{Zhao2020ApJS,Zhao2021A&A_detection}, we calculate the normalized cross helicity and residual energy to examine the Alfvénicity of the structures. The normalized cross helicity $\sigma_c$ and residual energy $\sigma_r$ were calculated from the Elsässer variables $z^{\pm} = \mathbf{u} \pm \delta \mathbf{b}$ with $\delta \mathbf{b} = \mathbf{B}/\sqrt{4 \pi n_p m_p}$, where $\mathbf{u}$ is the fluctuating velocity field, $\delta \mathbf{b}$ is the fluctuating magnetic field, $n_p$ is the proton number density, and $m_p$ is the proton mass:

\begin{equation}
\sigma_c(s, t) = \frac{\langle z^{+2} \rangle - \langle z^{-2} \rangle}{\langle z^{+2} \rangle + \langle z^{-2} \rangle} = \frac{2 \langle \tilde{\mathbf{u}} \cdot \tilde{\mathbf{b}} \rangle}{\langle \tilde{u}^2 \rangle + \langle \tilde{b}^2 \rangle},
\end{equation}
and
\begin{equation}
\sigma_r(s, t) = \frac{2 \langle z^{+} \cdot z^{-} \rangle}{\langle z^{+2} \rangle + \langle z^{-2} \rangle} = \frac{\langle \tilde{u}^2 \rangle - \langle \tilde{b}^2 \rangle}{\langle \tilde{u}^2 \rangle + \langle \tilde{b}^2 \rangle},
\end{equation}
where $z^{+}$ ($z^{-}$) represents the forward (backward) propagating modes with respect to the mean magnetic field orientation, and $\langle z^{+2} \rangle$ and $\langle z^{-2} \rangle$ represent the energy density in forward and backward-propagating modes, respectively. Alfvénic fluctuations are characterized by a high cross helicity ($|\sigma_c| \sim 1$), indicating dominant energy in either the $z^{+}$ or $z^{-}$ modes, and a low residual energy ($\sigma_r \sim 0$), which reflects a balance between kinetic and magnetic energies. In contrast, magnetic energy typically dominates over kinetic energy in SFRs, leading to a negative $\sigma_r$.

\begin{figure*}[h!]
\sidecaption
\centering
\makebox[\hsize][c]{\resizebox{1.0\hsize}{!}{\includegraphics{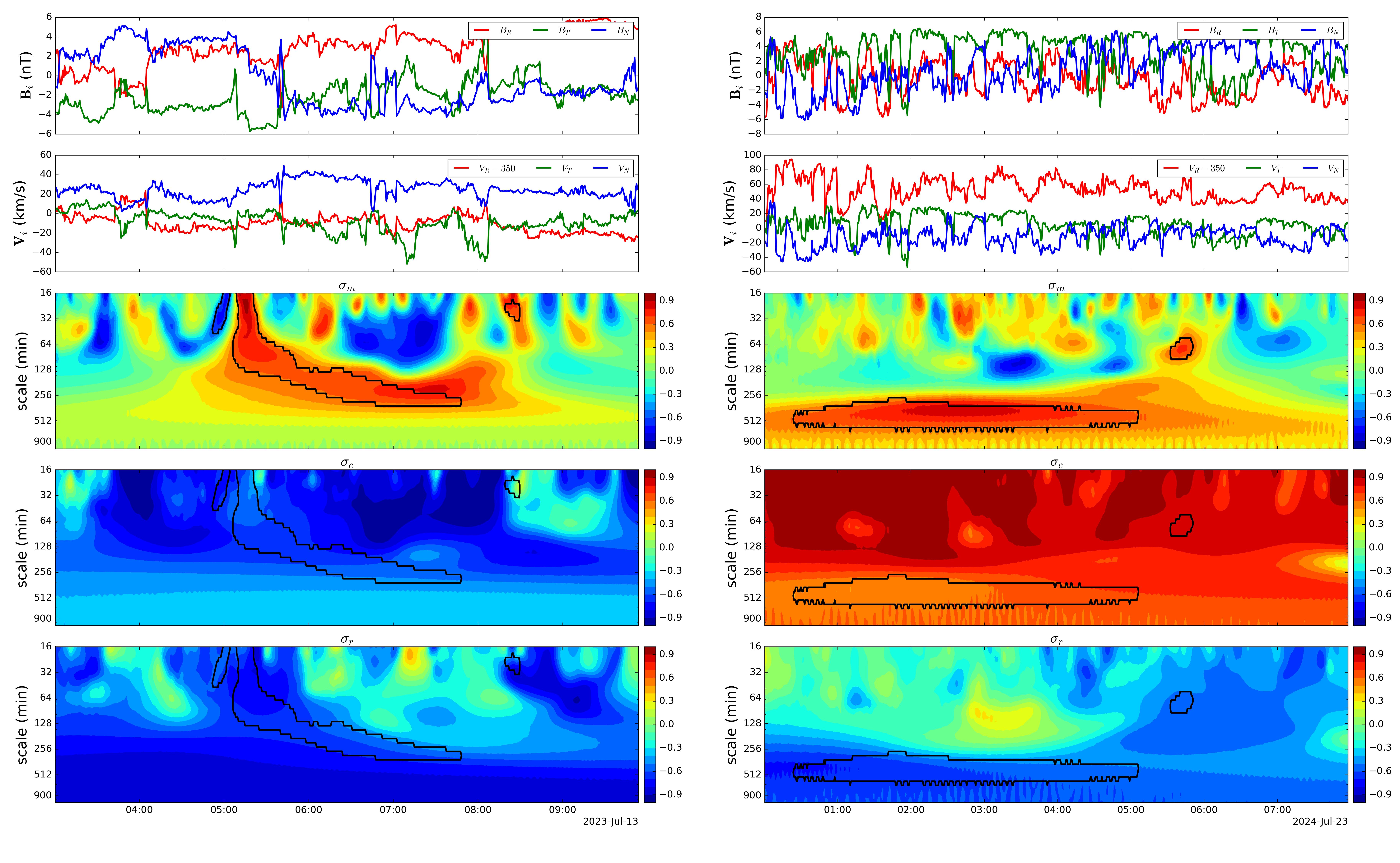}}}
\caption{ Wavelet spectrograms for events 1 (left) and 2 (right), corresponding to the onset phase of the SEP events. Top two panels: Magnetic field and velocity components. Bottom three panels: Wavelet spectrograms of the normalized magnetic helicity ($\sigma_{m}$), normalized cross helicity ($\sigma_{c}$), and normalized residual energy ($\sigma_{r}$). Black contours indicate magnetic flux ropes that satisfy the criteria $|\sigma_{m}| \geq 0.6$ and $\sigma_{r} \leq -0.5$. }
\label{fig:wavelet}
\end{figure*}

Figure~\ref{fig:wavelet} shows a zoom-in view of wavelet spectrograms for events 1 and 2. The period shown corresponds only to the onset phase of the SEP events. The top two panels display the magnetic field and velocity components. The bottom three panels present the wavelet spectrograms of the normalized magnetic helicity $\sigma_m$, the normalized cross helicity $\sigma_c$, and the normalized residual energy $\sigma_r$. We chose selection criteria ($|\sigma_m| \geq 0.6$ and $\sigma_r \leq -0.5$) to identify the possible SFRs,  indicated by the black contours.  These structures are related to high values of $|\sigma_c|$, which indicates that the structures are Alfvénic. For event 1, the largest flux rope displayed in the left panels persists from approximately 05:00 to 08:00 UT. For event 2, the largest flux rope shown in the right panels lasts from about 00:00 to 05:00 UT. Taken together, the selection criteria and high $|\sigma_c|$ values suggest that these structures might be Alfvénic SFRs.

\section{Estimation of the central axis length of magnetic flux ropes}
\label{appendix:sfr}

\begin{figure}
\sidecaption
\centering
\makebox[\hsize][c]{\resizebox{0.5\hsize}{!}{\includegraphics{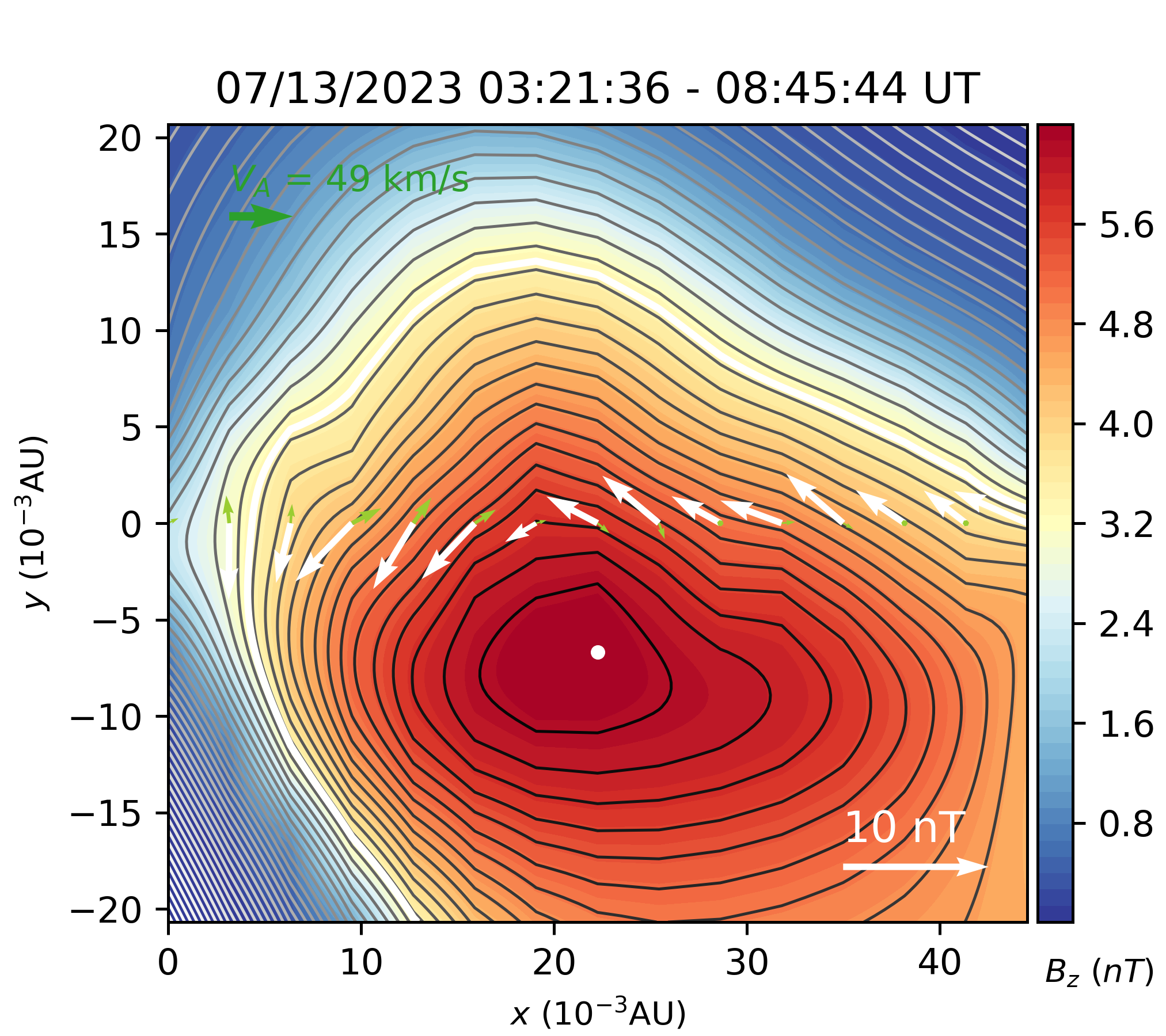}}}
    \caption{ Standard cross-section map from the GS reconstruction of the small-scale magnetic flux rope observed by Solar Orbiter on July 13, 2023, from 03:21:36 to 08:45:44~UT.  The colour map represents $Bz$ as indicated by the colour bar, while the black contours represent the transverse field lines. The spacecraft path is along the line $y=0$, with the measured transverse field vectors marked by the white arrows.   }
    \label{fig:GS_recon}
\end{figure}

Figure~\ref{fig:GS_recon} shows the GS reconstruction of the flux rope observed by Solar Orbiter between 03:21:36 and 08:45:44~UT on July 13, 2023, which is related to the onset of the bidirectional SEPs.  The colour scale represents the axial magnetic field component ($B_z$), while the white arrows indicate the direction and relative strength of the transverse field components ($B_t$). The reconstructed structure displays a well-defined helical magnetic topology, with a central core of enhanced $B_z$ and rotation of $B_t$ about the axis.

To constrain the global geometry of the flux rope, particularly its central axial length, we combined the path length inferred from VDA with the helical configuration revealed by GS reconstruction. Although the degree of helical winding may vary across the flux rope, we approximated the overall twist by using the mean value of the ratio $B_t / B_z$ along the spacecraft trajectory. This ratio quantifies the average helical winding of magnetic field lines about the central axis. We further assumed that the effect of axis curvature on the total path length can be neglected, and the magnetic field exhibits approximate cylindrical symmetry. The helical path length $L$ is then estimated as
\begin{equation}
L = \int_0^{L_0} \sqrt{1 + \left(\frac{B_t}{B_z}\right)^2} dl \approx \sqrt{1 + \left(\frac{B_t}{B_z} \right)^2 }L_0,
\end{equation}
where $L_0$ denotes the length of the central axis of the flux rope.  With $L$ determined from VDA and  $B_t / B_z$  from GS reconstruction, this relationship enables us to estimate $L_0$. For  event~1, as shown in Fig.~\ref{fig:vda}, the path length of anti-sunward streaming particles is approximately $1.75~\mathrm{au}$, while that of sunward streaming particles is about $3.04~\mathrm{au}$. Using the mean value $\langle B_t / B_z \rangle \approx 1.03$ derived from Fig.~\ref{fig:GS_recon}, these values yield $L_{0,\mathrm{sun}} \approx 1.19~\mathrm{au}$ for the connection to the sunward telescope and $L_{0,\mathrm{asun}} \approx 2.1~\mathrm{au}$ for the anti-sunward connection, respectively. The $L_{0,\mathrm{sun}}$ value closely matches the nominal Parker spiral length at $0.95~\mathrm{au}$ for a solar wind speed of $300~\rm{km/s}$, whereas $L_{0,\mathrm{asun}}$ places a constraint on the global expansion of the flux rope. However, due to the lack of information regarding the precise position of the flux rope apex, we are unable to fully determine the angular width of the structure. Nevertheless, given a total central axis length of $3.3$~au, the angular width of the flux rope is expected to be less than around $60^\circ$. If we assume that both legs below $0.95$~au are approximately $1.19$~au in length, the leading segment would be about $0.91$~au. Approximating this segment as a circular arc, this yields an angular width of roughly $55^\circ$. Since the actual apex height and expansion of flux rope remain uncertain, a complete understanding of the global topology of the flux rope is not possible from single-point measurements alone. Therefore, coordinated multi-spacecraft observations are crucial for fully constraining the overall structure.    

\end{appendix}

\end{document}